\documentclass[twocolumn,showpacs,preprintnumbers,amsmath,amssymb]{revtex4}

\begin{document}
\title{Almost-stationary motions and gauge conditions in General Relativity}

\author{C.~Bona$^1$, J.~Carot$^1$ and C.~Palenzuela-Luque$^2$}
\affiliation{ $^1$ Departament de Fisica,
Universitat de les Illes
Balears, Palma de Mallorca, Spain \\
$^2$ Department of Physics and Astronomy, Louisiana State
University, Louisiana, USA } \vspace{-1cm}

\begin{abstract}
An almost-stationary gauge condition is proposed with a view to
Numerical Relativity applications. The time lines are defined as
the integral curves of the timelike solutions of the harmonic
almost-Killing equation. This vector equation is derived from a
variational principle, by minimizing the deviations from isometry.
The corresponding almost-stationary gauge condition allows us to
put the field equations in hyperbolic form, both in the
free-evolution ADM and in the Z4 formalisms.
\end{abstract}


\maketitle

\section{Introduction}

It is well known that there is no preferred set of coordinate
frames in Einstein's theory of gravitation, which is known as
General Relativity precisely because of this fact. In practical
applications, however, one is forced to consider specific values
of the gravitational field components (spacetime metric, curvature
tensor) and this can be done only after choosing a specific
coordinate system. One just expects that this gauge choice will
not hide the physics of the problem behind a mask of non-trivial
coordinate effects.

Harmonic coordinate systems have deserved much interest since the
very beginning of Einstein's theory~\cite{DeDo21, Fock59}. The
coordinates themselves are defined by a set of four harmonic
spacetime functions $\{~\Phi^\mu\}$~, that is
\begin{equation}\label{full_harmonic}
 \square~ \Phi^{\mu} = 0~,
\end{equation}
where the box stands for the wave operator acting on functions. In
an harmonic coordinate system one takes $x^\mu \equiv \Phi^\mu$,
so that Einstein's field equations
\begin{equation}\label{Einstein}
  G_{\mu \nu} = 8~\pi~T_{\mu \nu}
\end{equation}
get a very convenient form, in which the principal part is simply
a scalar wave equation for every metric component~\cite{DeDo21,
Fock59}. This fact has been used by Yvonne Choquet-Bruhat for
proving the well-posedness of the Cauchy problem in General
Relativity~\cite{Choquet62}.

In the recent years, this simplification is being used for
building up Numerical Relativity codes, with interesting related
developments also on the theoretical side~\cite{Garfinkle, SSW02,
SW03, BSW04}. On the applications side, the 'gauge sources'
variant of the harmonic gauge~\cite{Friedrich96} has been used by
Pretorius in a numerical simulation of the evolution of a binary
black-hole system: a full quasi-circular orbit has been
achieved~\cite{Pretorius05a, Pretorius05b}. In this simulation,
however, a grid velocity has been introduced in order to get into
the co-rotating frame. This means that computational nodes are
rotating with respect to the coordinate system. In other words,
harmonic coordinates by themselves are not following the overall
rotation pattern of the black-hole binary system.

Binary systems provide a good example of almost-stationary
configurations. Take for instance the well known pulsar 1913+16:
it would be a perfect clock if we could just neglect the (very
small) energy loss due to gravitational radiation, getting then a
sort of steady system. One would like to choose a co-rotating
frame in order to get a clean view of symmetry deviations. This
case is representative of many other situations in which there is
not a exact symmetry. In these cases, the idea of approximate
symmetry, or that of almost-Killing vectors, would be of great
help in Numerical Relativity applications.

A precise implementation of the concept of almost-symmetry has
been provided by Matzner~\cite{Matzner68}. Starting from a
variational principle, it defines a measure of the symmetry
deviation of any given spacetime. This idea has been applied by
Isaacson~\cite{Isaacson68} to the study of high-frequency
gravitational waves, by defining a steady coordinate system in
which the radiation effects can be easily separated from the
background metric. More recently~\cite{Zala}, the same measure has
been considered as an inhomogeneity index of the spacetime, which
can be related with some entropy concept.

We are not interested here, however, in studying the spacetime
properties or in comparing different spacetimes. We will focus
instead in characterizing motions in arbitrary spacetimes. By a
motion we mean a congruence of time lines, that can then be
associated to the world lines of a system of observers. For
instance, geodesic motions (associated with freely falling
observers) or harmonic motions (associated with the observers at
rest in harmonic coordinate systems). Our goal is then to
characterize the motions that correspond to the physical idea of
almost-symmetry and to study the adapted coordinate systems with a
view to Numerical Relativity applications.

In this sense, we will see that our approach is more directly
related with the almost-Killing equation. This is a generalization
of the Yano-Bochner equation~\cite{Yano53}
\begin{equation}\label{Yano}
    \nabla_\nu~[~\xi^{\mu\,;\,\nu} + \xi^{\nu ;\,\mu}~] = 0~,
\end{equation}
which has been considered by York~\cite{York74} and others in
order to identify physically meaningful tensor components in
asymptotically flat spacetimes (transverse-traceless
decomposition). The same idea has also been applied to the
identification of asymptotic Killing vectors in Kerr
spacetime~\cite{Taubes78}.

These Killing-like equations are briefly reviewed in the next
section, where the almost-Killing equation is derived from a
variational approach. The main new results start however in the
third section, where harmonic almost-Killing motions are shown to
provide a convenient generalization of the standard harmonic
motions. This generalization is implemented through a true vector
condition, in contrast with the standard harmonic case, in which
one deals instead with the set of four scalar conditions
(\ref{full_harmonic}). This point is illustrated by considering
spherically symmetric spacetimes, where spherical coordinates are
incompatible with the standard harmonic condition but perfectly
allowed by the proposed generalization.

In section four, we consider the related problem of using harmonic
almost-stationary motions as gauge conditions, with a view to
Numerical Relativity applications. The adapted coordinate system
is used in order to show the hyperbolicity of the full system:
Einstein's field equations plus gauge conditions. Of course, any
hyperbolicity proof requires a specific formulation for the field
equations. We have chosen here the Z4 formalism~\cite{Z4} just for
simplicity, although the proposed almost-stationary gauge
condition should also work out with other hyperbolic formalisms.
The specific implementation we provide could be used as a guide
for any other particular choices. In order to illustrate this
point, we also show the hyperbolicity of the standard ADM
free-evolution approach when supplemented with the proposed
almost-stationary gauge condition.

\section{Almost-Killing vector fields}

\subsection{Killing-like equations}

Killing vectors can be defined as the solutions of the Killing
equation:
\begin{equation}\label{Killing}
    \mathcal{L}_\xi~(g_{\mu\nu}) =
    \nabla_\mu~\xi_\nu + \nabla_\nu~\xi_\mu = 0~.
\end{equation}
Their physical meaning can be better understood by considering an
adapted coordinate system. In the timelike case, for instance, we
can choose the time lines to be the integral curves of $\xi$ and
the time coordinate to be the special choice of the affine
parameter on these curves such that
\begin{equation}\label{adapted}
    \xi = \partial_t~.
\end{equation}
Then, the Killing equation (\ref{Killing}) reads simply
\begin{equation}\label{Killing_ad}
    \partial_t~g_{\mu\nu} = 0~,
\end{equation}
meaning that the metric is stationary, so that spacetime geometry
is preserved along the integral curves of $\xi$.

A well known generalization of the Killing equation
(\ref{Killing}) is given by the Affine Killing vectors (AKV),
namely the solutions of
\begin{equation}\label{AKV}
    \nabla_\rho~[~\mathcal{L}_\xi~(g_{\mu\nu})~] = 0~.
\end{equation}
The physical meaning is again more transparent if we express it
the adapted coordinate system (\ref{adapted}). Then, equation
(\ref{AKV}) amounts to
\begin{equation}\label{AKV_ad}
    \partial_t~\Gamma^\mu_{~\rho\sigma} = 0~.
\end{equation}
This means that the affine structure of the spacetime, given by
the connection coefficients $\Gamma^\mu_{~\rho\sigma}~$, is
preserved along the integral curves of $\xi$.

A interesting subset of AKV is that of the Homothetic Killing
vectors, defined as the solutions of
\begin{equation}\label{HKV}
    \mathcal{L}_\xi~(g_{\mu\nu}) = 2~g_{\mu\nu}
\end{equation}
(the factor two in the right-hand-side can be changed to any
non-zero value by a suitable rescaling of $\xi$). The physical
relevance of the Homothetic Killing vectors comes from the
invariance of equation (\ref{HKV}) under a rescaling of the
metric. This translates the idea that there is no preferred length
(time) scale in the spacetime, allowing scale-invariant
(self-similar) processes to develop. These processes have been
seen to arise in connection with critical phenomena in General
Relativity~\cite{Choptuik, Gundlach03}.

\subsection{A variational-principle approach: almost-Killing
vectors}

We will consider here a further generalization of (\ref{Killing}),
the almost-Killing equation (AKE) given by~\cite{Taubes78}
\begin{equation}\label{AKE}
    \nabla_\nu~[~\xi^{(\mu\,;\,\nu)} -
    \frac{\lambda}{2}~(\nabla\cdot\xi)~g^{\mu\nu}~] = 0~,
\end{equation}
where the round brackets denote symmetrization. The solution space
includes Killing vectors and AKV for any value of the constant
$\lambda$. The simplest parameter choice ($\lambda=0$),
corresponds to the Yano-Bochner equation~\cite{Yano53}. The case
$\lambda=\frac{1}{2}~$ corresponds to the Conformal AKE, which
includes conformal Killing vectors ($\lambda=\frac{\,2}{\,3}~$ for
a three-dimensional manifold, as in ref.~\cite{York74}\,).

The term 'almost-Killing' is justified by the fact that the AKE
equation (\ref{AKE}) can be obtained from a standard variational
principle
\begin{equation}\label{action}
    \delta S = 0~, \qquad S \equiv \int L ~\sqrt{g} ~d^4 x~\,,
\end{equation}
where the Lagrangian density $L$ is given by
\begin{equation}\label{Lag1}
    L = \xi_{(\rho\,;\,\sigma)}\xi^{(\rho\,;\,\sigma)} -
    \frac{\lambda}{2}~(\nabla\cdot\xi)^2~,
\end{equation}
and the variations of the field $\xi$ are considered in a fixed
spacetime. The covariant conservation law (\ref{AKE}) provides
then a precise meaning to the heuristic concept of approximate
Killing vectors. This was not obvious a priori, because the
Lagrangian (\ref{Lag1}) is not a positive-definite quantity: the
outcome of the minimization process was not granted to include the
zeros of (\ref{Lag1}).

The mathematical structure of the AKE is more transparent if we
rewrite it in the equivalent form
\begin{equation}\label{Eqs2}
    \square \xi_\mu + R_{\mu\nu}\,\xi^\nu
    + (1-\lambda)~\nabla_\mu (\nabla\cdot\xi) = 0~,
\end{equation}
where we have just reversed the order of some covariant
derivatives in (\ref{AKE}). This 'wave equation' form of the AKE
can be alternatively obtained from the Lagrangian
\begin{equation}\label{Lag2}
   L' = \xi_{\rho\,;\,\sigma}\xi^{\rho\,;\,\sigma} -
   R_{\rho\sigma}\,\xi^\rho\xi^\sigma +
    (1-\lambda)~(\nabla\cdot\xi)^2~,
\end{equation}
which is of course equivalent to the original one, modulo a
four-divergence:
\begin{equation}\label{equivL}
    L' = 2L - \nabla_\sigma ~[~(\xi\cdot\nabla)~\xi^\sigma -
    \xi^\sigma~(\nabla\cdot\xi)~]~.
\end{equation}

The principal symbol of the differential operator in either form
of the AKE can be written in Fourier space as
 \begin{equation}\label{symbol}
    k^2~\delta^\mu_{~\nu} + (1-\lambda)~k^\mu k_\nu~,
\end{equation}
so that:
\begin{itemize}
    \item The characteristic hypersurfaces are the light cones
    ($k^2=0$).
    \item The symbol (\ref{symbol}) is singular for $\lambda=2$. For
    vacuum spacetimes, this case corresponds to Maxwell's
    equations for the electromagnetic potential~\cite{Taubes78}. A
    supplementary condition (like the 'Lorentz condition'
    $\nabla\cdot\xi=0$) would be then
    required in order to get a unique solution.
    \item On non-characteristic hypersurfaces, the symbol
    (\ref{symbol}) can be algebraically inverted for $\lambda \neq 2$.
\end{itemize}
The last point is a strong indication of the existence of
solutions for $\xi$ in any given spacetime, for every set of
non-characteristic initial data. Of course, this is not a $100$\%
rigorous proof because the straightforward passage to Fourier
space ignores the coordinate dependence of the metric (the
standard 'frozen coefficients' approach). But this gives us a
sound basis for assuming in what follows that solutions for $\xi$
may be constructed in any given spacetime.

To be more specific, the initial value problem can be expressed as
follows
\begin{itemize}
    \item We can freely choose the values of $\xi$ on a given
initial hypersurface (let us say $t=0$). The space derivatives of
$\xi$ can then be computed from these values.
    \item The time derivative of $\xi$ can also be freely
specified on the initial hypersurface. In this way, we have the
full set of first covariant derivatives $\xi_{\mu;\nu}$ at $t=0$.
    \item In order to propagate these values along the time lines,
we must compute the second time derivative of $\xi$ from the
second order equation (\ref{Eqs2}). This can always be done for
the space components $\xi_i$ provided that $g^{00} \neq 0$,
meaning that the initial hypersurface is not tangent to the local
light cone. In the case of the $\xi_0$ component, we must require
in addition that $\lambda \neq 2$, as it was to be expected from
the results of the preceding paragraph.
\end{itemize}

In the timelike case, the integral curves of $\xi$ can be
interpreted as the world-lines of a set of observers. As far as
there is one solution for every set of (non-characteristic)
initial data, we can interpret the set of solutions as providing a
set of motions that minimize the deviation from isometry along the
congruence of time lines. This justifies the name of
'almost-stationary motions' for the timelike solutions of the AKE.

\section{Harmonic almost-stationary motions}

We will consider now the particular parameter choice $\lambda=1$,
in which the principal part of equation (\ref{Eqs2}) is harmonic,
that is
\begin{equation}\label{Eqs2H}
    \square \xi^\mu + R^\mu_{~\nu}\,\xi^\nu = 0~,
\end{equation}
so that the resulting timelike solutions will be called 'harmonic
almost-Killing motions'. The conservation-law version (\ref{AKE})
can then we written as
\begin{equation}\label{HAKE}
    \nabla_\nu~[~\frac{1}{\sqrt{g}}~~\mathcal{L}_\xi~(
    \sqrt{g}~g^{\mu\nu})~] = 0~.
\end{equation}
Notice that the Lagrangian $L'$ in this case, namely
\begin{equation}\label{Lag21}
   L' = \xi_{\rho\,;\,\sigma}\xi^{\rho\,;\,\sigma} -
   R_{\rho\sigma}\,\xi^\rho\xi^\sigma ~,
\end{equation}
gets an interesting 'kinetic minus gravitational' form.

Both forms (\ref{Eqs2H}) and (\ref{HAKE}) of the Harmonic AKE
equation (HAKE) suggest a close relationship with harmonic
coordinates. This relationship is again more transparent in the
adapted coordinate system, where (\ref{HAKE}) leads to
\begin{equation}\label{Gamma}
    g^{\rho\sigma}~\partial_t~ \Gamma^\mu_{~\rho\sigma} = 0~
\end{equation}
(compare with Eq.~\ref{AKV_ad} for AKV), whereas the harmonic
coordinates condition (\ref{full_harmonic}) reads just
\begin{equation}\label{harmonic}
    \Gamma^\mu \equiv g^{\rho\sigma}~\Gamma^\mu_{~\rho\sigma} = 0~
\end{equation}
in adapted coordinates.  One can then write (\ref{Gamma}) as
\begin{equation}\label{Gamma2}
    \partial_t~\Gamma^\mu =
    \Gamma^\mu_{~\rho\sigma}~\partial_t~ (g^{\rho\sigma})~,
\end{equation}
so that it is clear that the flow associated with harmonic
coordinates will provide a first integral for the HAKE in the weak
field limit (where only linear terms are retained). This fact can
be relevant for the characterization of the gravitational waves
degrees of freedom in asymptotically flat
spacetimes~\cite{Isaacson68}.

Note also that the HAKE (\ref{HAKE}) is a true vector equation,
whereas the standard harmonic condition (\ref{full_harmonic}) is
rather a set of four scalar equations. This difference is
important, because the congruence of time lines in a given motion
is defined by its tangent vector field. One can, for instance,
relabel the particular time lines by an arbitrary time-independent
coordinate transformation, namely
\begin{equation}\label{transf}
    x^i = f^i(y^j) \qquad  i,j=1,2,3~,
\end{equation}
while keeping the same expression (\ref{adapted}) for the timelike
tangent vector. The transformation (\ref{transf}) allows one to
select the type of space coordinate system (cylindrical,
spherical, or whatsoever) which is more adapted to any specific
problem. In numerical simulations, the transformation
(\ref{transf}) corresponds to the freedom of choosing an arbitrary
space coordinate system on the initial hypersurface.

This is not the case for the standard harmonic coordinates choice.
In order to illustrate this, we will consider for instance a
spherically symmetric line element, namely
\begin{equation}\label{spheric}
    ds^2 = -\alpha^2~dt^2 + X^2~dr^2 +
    Y^2~[d\theta^2+sin^2(\theta)~d\varphi^2]~,
\end{equation}
where all the metric functions ($\alpha$, $X$, $Y$) depend only on
($t$, $r$). In this case, the time and radial components
components of the HAKE (\ref{Gamma}) provide conditions for the
corresponding ($t$, $r$) coordinates, whereas the angular
components are identically satisfied in the adapted coordinate
system form (\ref{Gamma}). In the standard harmonic case, however,
one gets
\begin{equation}\label{harmkk}
    \Gamma^\varphi = 0~,\qquad
    \Gamma^\theta=-\frac{cot(\theta)}{Y^2} \neq 0~,
\end{equation}
so that spherical coordinates happen to be incompatible with the
harmonic condition (\ref{harmonic}).

This lack of versatility  of the standard harmonic coordinates can
be a serious drawback in Numerical Relativity applications, where
one could be unable to fully adapt the coordinate frame to the
features of the physical system under consideration. We hope that
the proposed almost-stationary generalization (\ref{Gamma}) will
contribute to avoid this complication.

\section{Almost-stationary gauge conditions}

Standard harmonic motions, as defined by (\ref{harmonic}), have
been used recently in advanced Numerical Relativity
applications~\cite{Garfinkle, SSW02, SW03, BSW04, Pretorius05a,
Pretorius05b}. Note however that in this context one is building
up the coordinate system and the spacetime itself at the same
time, whereas the spacetime was supposed to be given in the
previous sections. To be more precise, Einstein's field equations
(\ref{Einstein}) must be coupled with the gauge condition. In the
harmonic case, this condition is given by eq.~(\ref{harmonic}).

The principal part of Einstein's equations can be written in the
DeDonder-Fock form~\cite{DeDo21,Fock59}, namely
\begin{equation}\label{DeDonder}
 \square~ g_{\mu \nu} - \partial_{\mu} \Gamma_{\nu}
  - \partial_{\nu} \Gamma_{\mu}  = \cdots
\end{equation}
so that, by choosing the spacetime coordinates to be the solutions
of (\ref{full_harmonic}), one ensures the vanishing of
$\Gamma^\mu$ and the field equations can be relaxed to a system
whose principal part consists in a scalar wave equation for every
metric component, namely
\begin{equation}\label{boxg}
 \square~ g_{\mu \nu} = ...
\end{equation}

Note that the metric in (\ref{harmonic}, \ref{boxg}) is
over-determined, because one gets in all $14$ equations for only
$10$ metric components. This can be better understood by
introducing a 'zero vector' $Z^\mu$ as an additional dynamical
field (Z4 system~\cite{Z4}), so that the field equations read
\begin{equation}\label{Einstein_extended}
  R_{\mu \nu} + \nabla_{\mu} Z_{\nu} + \nabla_{\nu} Z_{\mu} =
  8~ \pi~ (T_{\mu \nu} - \frac{1}{2}~T~ g_{\mu \nu})~.
\end{equation}

The principal part of the Z4 field equations in the DeDonder-Fock
form reads now
\begin{equation}\label{DeDonder_ext}
 \square~ g_{\mu \nu} - \partial_{\mu} (\Gamma_{\nu} + 2~ Z_{\nu})
   - \partial_{\nu} (\Gamma_{\mu} + 2~ Z_{\mu}) = \cdots~,
\end{equation}
so that the relaxed system (\ref{boxg}) is recovered by setting
the values of $Z^\mu$ to be
\begin{equation}\label{harmonic extended}
 Z^\mu = - \frac{1}{2}~\Gamma^{\mu},
\end{equation}
which can be considered just an extension of the harmonic
coordinates condition (\ref{full_harmonic}), namely
\begin{equation}\label{full_harmonic extended}
 \square~ \Phi^{\mu} = 2~Z^\mu~.
\end{equation}

As far as one recovers in this way exactly the same relaxed system
(\ref{boxg}), one gets exactly the same solutions for the metric.
The extra quantities $Z^\mu$ just allow us to monitor to which
extent the harmonic coordinates condition (\ref{harmonic
extended}), when considered as a constraint on the computed
metric, is actually verified. In this sense, it is useful to
consider the four-divergence of the Z4 field equations
(\ref{Einstein_extended}), namely
\begin{equation}\label{laplacian}
  \square~ Z_{\mu} + R_{\mu \nu}~ Z^{\nu} = 0~,
\end{equation}
which can be interpreted as the constraint-propagation law. It
follows that the constraint-violation vector $Z^\mu$ obeys
precisely the HAKE (\ref{Eqs2H}).

Note that the Z4 system has been used here just as a convenient
analysis tool in order to discuss the overlap between the $10$
equations of the relaxed system and the $4$ equations of the
harmonic coordinates condition. The same overlap will occur when
replacing the standard harmonic coordinates condition
(\ref{harmonic}) by the HAKE (\ref{Gamma}). This suggests to
consider, in the Z4 context, the analogous replacement of
(\ref{harmonic extended}) by
\begin{equation}\label{Gamma extended}
    g^{\rho\sigma}~\partial_t~ (\Gamma^\mu_{~\rho\sigma})
    + 2~\partial_t~Z^\mu = 0~.
\end{equation}
A covariant expression for (\ref{Gamma extended}) is given by
\begin{equation}\label{HAKEZ4}
    \nabla_\nu~[~\frac{1}{\sqrt{g}}~~\mathcal{L}_\xi~(
    \sqrt{g}~g^{\mu\nu})~] =
    2~\mathcal{L}_\xi~(Z^\mu)~.
\end{equation}

The principal part of (\ref{Gamma extended}) can be written now
simply as
\begin{equation}\label{HAKE prin}
    \partial_t~ (\Gamma^\mu + 2~Z^\mu) = \cdots~,
\end{equation}
so that (the principal part of) the full system
(\ref{DeDonder_ext}, \ref{HAKE prin}) gets a triangular form. The
characteristic lines can easily be identified:
\begin{itemize}
    \item the time lines, as it follows from the fact that the
    'Gamma sector' equations (\ref{HAKE prin}) are yet in diagonal form
    \item the light rays, as it follows from the fact that the
    only non-diagonal terms in the 'metric sector' equations
    (\ref{DeDonder_ext}) are just coupling terms with the Gamma
    sector, which is itself in diagonal form.
\end{itemize}

We can conclude that the full differential system formed by the Z4
system (\ref{Einstein_extended}) and the (extended) HAKE condition
(\ref{Gamma extended}) is hyperbolic. The only trouble can arise
at the specific points where the harmonic almost-stationary vector
$\xi$ is light-like ($\xi^2=0$), so that the corresponding time
line gets tangent to the light cone. The non-diagonal coupling
terms in (\ref{DeDonder_ext}) would then prevent the full
diagonalization of the characteristic matrix at this specific
point. This would be for instance the case of the apparent horizon
in stationary spacetimes, when the Killing vector is selected as a
solution of the HAKE equation.

Let us note again that we are using here the Z4 formalism just as
an analysis tool, which allows us to monitor the evolution of the
constraint violations. The proposed coordinate conditions
(\ref{Gamma}) are actually independent of the hyperbolic formalism
one likes to choose for the field equations. In order to
illustrate this point, let us take for instance the standard ADM
free-evolution approach:
\begin{itemize}
    \item The original field equations (\ref{DeDonder}) are
considered. However, only the space components
\begin{equation}\label{ADM}
 \square~ g_{ij} - \partial_{i} \Gamma_{j}
  - \partial_{j} \Gamma_{i}  = \cdots
\end{equation}
are kept, because the four remaining combinations provide just
constraints which are not solved in the free evolution approach.
    \item The almost-stationary conditions (\ref{Gamma}), with
principal part
\begin{equation}\label{Gamma0i}
    \partial_t \Gamma^{0} = \cdots \qquad
    \partial_t \Gamma_{i} = \cdots
\end{equation}
which provide the missing evolution equations for the remaining
metric components $g^{00}$, $g_{0i}$, respectively.
\end{itemize}
We can see by inspection that the principal part of the full
system (\ref{ADM}, \ref{Gamma0i}) is also in triangular form. The
characteristic lines are again either the time lines ('Gamma
sector') and the light cones ('metric sector'). The resulting ADM
system, when supplemented with the almost-stationary gauge
condition, is then hyperbolic, provided that the time lines do not
get tangent to the light cones.

We can conclude that the same qualitative behavior is obtained
both in the Z4 and in the ADM frameworks when using the
almost-stationary gauge condition. We do not expect then any
essential difficulty in adapting this coordinate condition to
other hyperbolic formalisms which are being used in Numerical
Relativity applications.

{\em Acknowledgements: The authors are indebted with Luis Lehner
for providing useful suggestions and comments. This work has been
supported by the Spanish Ministry of Science and Education through
the research project number FPA2004-03666. }

\bibliographystyle{prsty}

\end{document}